# Photoluminescence of oxygen vacancies and hydroxyl group surface functionalized SnO$_2$ nanoparticles

Venkataramana Bonu,[*a] Arindam Das,[*a] S. Amirthapandian,[b] Sandip Dhara,[a] and Ashok Kumar Tyagi[a]

We report, for the first time, the luminescence property of the hydroxyl group surface functionalized quantum dots (QDs) and nanoparticles (NPs) of SnO$_2$ using low energy excitations of 2.54 eV (488 nm) and 2.42 eV (514.5 nm). This luminescence is in addition to generally observed luminescence from 'O' defects. The as-prepared SnO$_2$ quantum dots (QDs) are annealed at different temperatures in ambient conditions to create varied NPs in size. Subsequently, average size of the NPs is calculated from the acoustic vibrations observed at low frequencies in the Raman spectra and by the transmission electron microscopic measurements. Detailed photoluminescence studies with 3.815 eV (325 nm) excitation reveal the nature of in-plane and bridging 'O' vacancies as well as adsorption and desorption occurred at different annealing temperatures. X-ray photoelectron spectroscopy studies also support this observation. Defect level related to the surface –OH functional groups shows a broad luminescence peak around 1.96 eV in SnO$_2$ NPs which is elaborated with the help of temperature dependent studies.

## Introduction

SnO$_2$ is an *n*-type wide band gap (3.6 eV) semiconductor. The wide usage it in the areas such as gas sensing, transparent electrodes, catalyst, solar cell, Li-ion batteries made SnO$_2$ a technologically important material.[1] In all these important applications defects in SnO$_2$ play a vital role.[1-4] Prominent defects in SnO$_2$ crystals are 'O' vacancies, which create donor states showing *n*-type behavior.[1] Electrons from these donor states will participate in the formation of a depletion layer on the surfaces of SnO$_2$ crystals which is important for the gas detection process. In our earlier report in-plane 'O' (O$^P$) and bridging 'O' (O$^B$) vacancy defects were found to play crucial role in the detection of CH$_4$ gas at low temperatures.[5] Combination of shallow donor levels and wide band gap made this material transparent conductor.[6-9] Ferromagnetic nature in the SnO$_2$ nanostructures was also ascribed to the 'O' defects.[10-12] The above discussions emphasizes the importance of defects in SnO$_2$ and investigation of their properties. A broad photoluminescence (PL) peak observed for SnO$_2$ nanostructures around ~2 eV was ascribed to oxygen vacancy related defects.[12-15] After controled annealing of the amorphous SnO$_2$ thin films, sharp ultra violet light was found to emit which was utilized in fabricating diode with *p*-GaN.[16] Therefore improved understanding to gain the control over the oxygen defects in SnO$_2$ is technologically important.

Along with 'O' related defects, formation of hydroxyl (-OH) groups on the surface of SnO$_2$ creates another important point defects. Water can readily dissociate even on ideal oxide surfaces at room temperature.[17] In the case of nanoparticles (NPs), which possess a large number of surface defects, a high density of under-coordinated metal and oxygen atoms drastically increase chemisorptions of the polar molecule in the ambient conditions. These -OH groups on metal oxides considerably affect the electronic properties and surface chemistry of SnO$_2$.[18-21] For instance, formation of -OH groups on the metal oxides increases density of electrons in the conduction band.[18] It also influences optical property as shown by Sharma *et al.* for ZnO NPs. They observed a PL peak around 2.1 eV due to the -OH functionalization.[22] Using density functional theory (DFT) calculations, Valentin *et al.*[23] showed the electron trapping nature of -OH group present on the TiO$_2$ surfaces. Similarly, DFT calculations of -OH groups attached to Si surfaces revealed formation of donor defect states due to certain kind of Si-OH.[24] Based on various factors (*e.g.*, season, temperature, location) moisture in atmosphere can vary from ca. 0.1 up to 4 vol%, which notably changes the electronic properties of the sensors and influences their sensitivity.[24]

There are studies on the annealing effect on 'O' defects of the crystalline SnO$_2$ and desorption of 'O' with temperature.[14] In this context, there is lack of detailed investigation on the luminescence properties of the oxidation and desorption process of 'O' for quantum dots (QDs) and nanostructures of SnO$_2$. Additionally, there is scope of detailing luminescence study on the role of -OH groups that are ubiquitously present on the SnO$_2$ NPs surfaces.

Here we present luminescence properties of 'O' vacancy defects in the SnO$_2$ QDs and -OH functionalized SnO$_2$ NPs. A systematic study is conducted for the influence of annealing temperature on the size of the NPs. The sizes were calculated from the low frequency Raman scattering (LFRS) and was correlated to high resolution transmission electron microscopy (HRTEM) measurements. The O$^P$ and O$^B$ vacancies related luminescence were probed by the excitation energy 3.815 eV, whereas distinct defect levels created by -OH groups were elucidated by the low energy excitations of 2.54 and 2.42 eV preserving the pristine nature of the functionalized groups.

## Experimental Section

### Synthesis and Characterization

Detailed synthesis of SnO$_2$ QDs was described in our earlier study [3]. In brief, NH$_4$OH (MERCK) was added drop wise in to stannic chloride (SnCl$_4$, Alfa Aesar) under magnetic stirring at 80 °C. The resulted white gel was washed with Millipore water, and dried at 100 °C. This product is the as-prepared SnO$_2$ QDs and hereafter will be called as sample A (100 °C). The as-prepared pristine material was further annealed at 200, 250, 300, 500 and 800 °C for 1 h at ambient conditions in horizontal quartz tube furnace. The samples annealed at different temperatures will be termed as B (200 °C), C (250 °C), D (300 °C), E (500 °C) and F (800 °C).

HRTEM images were recorded with the aid of Libra 200 Zeiss electron microscope. X-ray diffraction (XRD) patterns (Bruker) were obtained using Cu–Kα (1.5406 A°) radiation. Micro-Raman spectroscopy (InVia, Reinshaw) was carried out using 514.5 nm excitation of an Ar+ laser with 1800 gr/mm grating, and thermo electric cooled CCD detector in the back scattering mode. The PL spectra (InVia, Reinshaw) were acquired using 514.5 nm (2.42 eV), 488 nm (2.54 eV) (excitation of an Ar+ laser) and 325 nm (3.815 eV) (He-Cd laser) as the excitation sources. Temperature dependent PL measurements were performed in adiabatic stage (Linkam, UK). Fourier transform infrared spectroscopy (FTIR, Bruker MB-3000) was used to probe infrared active vibrational modes.

## Results and Discussion

### Structural studies and morphological features

XRD patterns of the samples A (100 °C), D (300 °C) and F (800 °C) are shown in Fig. 1. All diffraction peaks of sample F are indexed following the rutile tetragonal crystalline phase of SnO$_2$ (JCPDS #41-1445). No other phase of SnO$_2$ could be observed. Sharp diffraction peaks in the sample F (800 °C) indicates presence of long range order in these SnO$_2$ NPs. The broad peaks are observed for the samples A and D at 2θ values of 26.5º, 33.8º, 51.7º, and 64.7º which closely match to planes (110), (101), (211) and (112) of rutile tetragonal SnO$_2$, respectively. This broadening of diffraction planes can be ascribed to the presence of short range order that is very small crystallites and also to the non uniform strain.[25] The mean crystallite size was calculated following Debye-Scherrer equation, R=0.89λ/βcosθ, where R = mean crystallite size, β = full width at half maximum (FWHM). Average sizes of the NPs found to be of 1.9±0.2 nm, 3.9±0.2 nm and 23±1 nm for the sample A (100 °C), D (300 °C) and F (800 °C), respectively. However in our earlier report[25] we have shown presence of non uniform strain in the sample A using Williamson-Hall (W-H) plot. After eliminating the non uniform strain contribution to FWHM value, the average size of the sample A was then found to be 2.3 ± 0.1 nm.

HRTEM image of D (300 °C) is shown in Fig. 2(a). Crystalline (110) plane with a *d*-spacing of 3.38 Å corresponding to rutile SnO$_2$ (JCPDS #41-1445) is shown as zoomed inset. Mean diameter is ~ 4.1±0.2 nm from the size distribution plot (Fig. 2(b)). TEM image of sample E (500 °C) is shown in Fig. 3(a). The average particle size grows by annealing to ~ 9±0.7 nm as shown in Fig. 3(b). HRTEM image in Fig. 3(c) confirms crystalline (110) plane with a *d*-spacing value of 3.36 Å (inset).

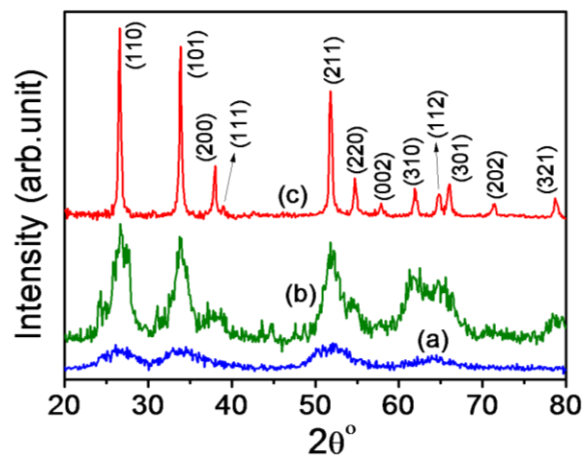

Fig. 1 XRD of samples (a) A (100 °C), (b) D (300 °C) and (c) F (800 °C)

Fig. 3(d) displays selected area electron diffraction (SAED) pattern. These ring like patterns indicates existance of crystallites in all possible orientations. Rings are indexed as (110), (101) and (211) planes which correspond to the rutile SnO$_2$ phase. Detailed structural characterization of the sample A (100 °C) and F (800 °C) were discussed in our erlier report.[3] Average size of the sample A and F were measured using HRTEM and found to be 2.4 and 25 nm, respectively.[3] Obtained NPs sizes from the TEM match well with XRD results.

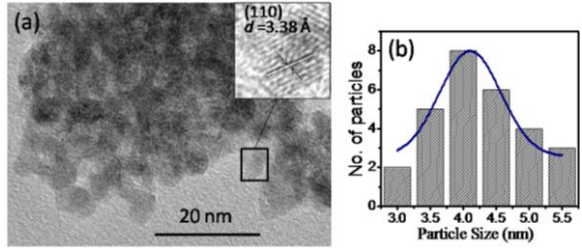

Fig. 2 (a) HRTEM image of SnO$_2$ NPs annealed at 300 °C. Inset shows crystalline (110) plane of rutile tetragonal SnO$_2$ (b) Gaussian fitting of size distribution curve.

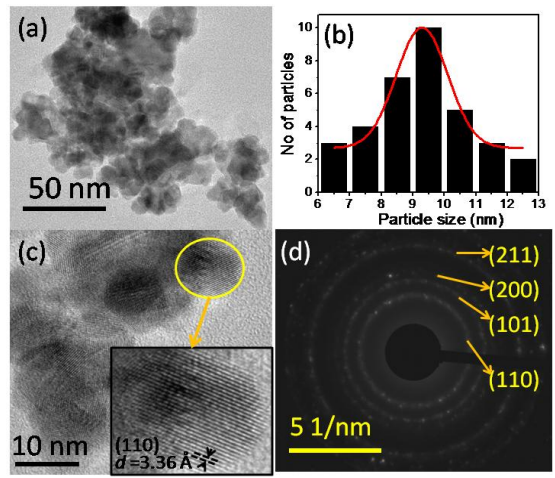

FIG. 3. (a) Low magnification TEM image of sample F (500 oC), (b) Particle size distribution (c) HRTEM image (d) SAED pattern of sample F (500 oC).

Raman spectra recorded for all the samples are shown in Fig. 4. Raman modes at 474, 633 and 775 cm$^{-1}$ of the annealed samples are assigned to $E_g$, $A_{1g}$ and $B_{2g}$ symmetries, respectively.[26] Peak at 575 cm$^{-1}$ (termed as $D$), uniquely found in ultra small SnO$_2$ NPs, is attributed to optically inactive $A_{2g}$ mode by following Matossi force constant model which considers the modified bond length, space symmetry reduction and lattice distortion due to 'O' vacancies in SnO$_2$ NPs.[27]

Low frequency Raman scattering (LFRS) from spherical acoustic vibrations of free SnO$_2$ NPs can be utilized for measuring the mean particle size of NPs.[28,29] Peak positions of low frequency mode for different particles sizes are shown in Fig. 4. There is observable red shift while the size of the particle increases with annealing temperatures. According to Lamb's theory, size of the NP relates to low frequency $w_{l,n}^{sph}$ as

$$w_{l,n}^{sph} = S_{l,n} \frac{v_l}{cR} \quad \ldots\ldots\ldots\ldots (1)$$

Where $v_l$ denoting the average transverse sound velocity of SnO$_2$ is 6530 m/s,[28] $R$ is the diameter of a NP, and $c$ is velocity of light in vacuum. Coefficient $S_{l,n} = 0.887$ strongly depends on the ratio of the transverse to longitudinal, $v_t/v_l$ sound velocities.[29] Average particle sizes 2.41 ± 0.15 nm, 2.66 ± 0.18 nm and 3.1 ± 0.22 nm were calculated using LFRS model for the samples A (100 °C), B (200 °C), and C (250 °C), respectively. The values are plotted in Fig. 5. Sizes calculated using TEM for samples D (300 °C), E (500 °C) and F (800 °C) are also indicated in Fig. 5. The sizes of NPs measured by different methodologies are tabulated in the supplementary information (TABLE SI). Due to the limitation of the edge filter in our Raman instrument, modes below 50 cm$^{-1}$ could not be probed accurately to calculate the average size of other samples.

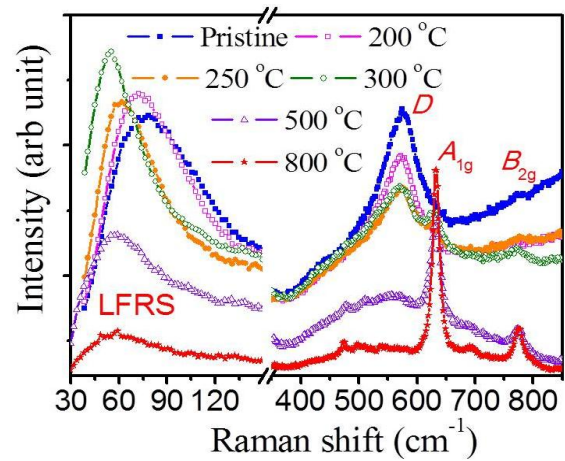

Fig. 4 Raman spectra of SnO$_2$ NPs annealed at different

As the particle radius goes below the Bohr radius, $r_B$, (2.7 nm for SnO$_2$ [30]) the electron hole motion is not correlated. In that case, the effective band gap energy is given by,

$$E_g^{eff} = E_g + \frac{\hbar^2 \pi^2}{2\mu r^2} - \frac{1.8 e^2}{\varepsilon r} + \cdots \quad \ldots\ldots\ldots\ldots (2)$$

If the NP size is higher than $r_B$, 3rd term (Coulomb interaction) in above equation is considered to be negligible. In the equation, $r$ is the particle radius, and μ is the effective reduced mass (0.27m$_e$ for SnO$_2$).[31] $E_g$ is the bulk band gap energy (3.6 eV) and the dielectric constant of SnO$_2$ is 14. Band gaps of the samples are calculated (supplementary TABLE S1) and plotted with varying NP sizes (Fig. 5). As size of the NPs increases, the corresponding band gap decreases. Size of the NPs does not change considerably until an annealing temperature of 300 °C with a size of 4.1 nm. Interestingly, NP size drastically increases to 9.5 nm upon annealing at a temperature of 500 °C. Thus it was inferred that the grain growth, which was the third stage of annealing process started effectively at 500 °C onwards, whereas other two initial steps of annealing process, recovery and re-crystallization could only take place predominantly until a temperature of 300 °C.

XPS analysis for NPs of sizes of 4, and 25 nm was carried out and reported earlier.[5] Atomic weight percentage ratios between 'O' and 'Sn' (O:Sn) were approximately 1.7 and 1.5 for 4 and 25 nm NPs, respectively. It indicates increasing oxygen deficiency on the surface of 25 nm NPs than that of in the 4 nm NPs. This enhanced defects density manifests as a result of strong desorption of 'O's from the surface of NPs due to annealing of the as-prepared material at high temperatures of 800 °C.

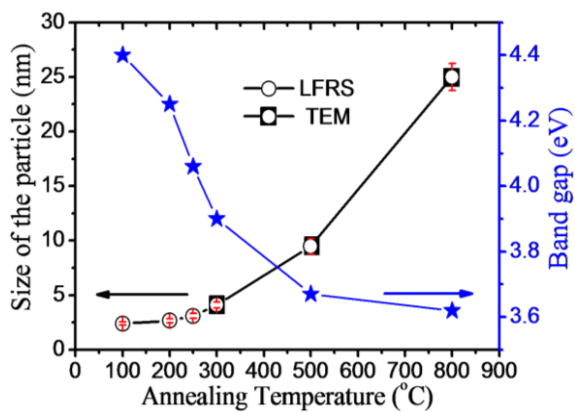

Fig. 5 Size and band gap of the $SnO_2$ NPs with respect to annealing temperature. Symbols ○ and □ represents the sizes determined by using LFRS and TEM studies, respectively. Error bar added for the NPs size values.

**Photoluminescence of 'O' vacancy related defects.**
PL studies are carried out using excitation energy of 3.815 eV in the ambient conditions. The excitation energy is found to be lower than the band gap of the samples A (100 °C), B (200 °C), C (250 °C) and D (300 °C) (Fig. 5), and hence defect states only are probed. Fig. 6 shows PL spectra of NPs, which are excited with 3.815 eV. A large PL intensity around 2 eV is observed for as-prepared QDs (A (100 °C)). Similar luminescence pattern is observed for $SnO_2$ nanostructures and it is attributed to `O` defects.[5,13,14,32-36] Notably, PL intensity around 2 eV (Fig. 6) decreases with the increasing annealing temperature and almost reduces to negligible value for the sample D (300 °C) of particle size 4.1 nm. It indicates that the defect centres responsible for strong luminescence are annihilated, possibly by oxidation during annealing in air atmosphere. However, PL intensity around 2 eV further increases for the samples E of particle size of 9 nm and the sample F of particle size of 25 nm. Here desorption of 'O' at high temperature incurs the 'O' defects which leads to further enhancement in PL intensity. This observation supports the XPS results. To get a better insight of the various defects, each plot is deconvoluted in to five Gaussian peaks being centred on around 1.8, 1.97, 2.12, 2.29 and 2.45 eV. PL spectra with Gaussian peak fittings are shown in the supplementary information Fig. S1(a-e). Peak positions and fractional areas of each peak are given in TABLE I. The fractional area is simply proportional to the fractional number of electronic transitions in these five channels. There are two types of 'O' vacancies in the $SnO_2$ crystal which are $O^B$ and $O^P$ vacancies (Fig. 7). Yellow luminescence (YL) (~2 eV) is attributed to the defect states created by $O^B$ vacancies [5,13,14] and the blue-green luminescence (BGL) is due to $O^P$ vacancies.[5,14] In Fig. S1, luminescence in the first three channels at 1.8, 1.97 and 2.12 eV are considered to be related to $O^B$ vacancies whereas luminescence centred at 2.29 and 2.45 eV channels are regarded as a result of $O^P$ vacancies. Detailed of the luminescence intensity from each of these two types are given in TABLE I. For sample A of particle size 2.4 nm, fraction of YL and BGL are 73.3 % and 26.7 %, respectively. After annealing the sample at 200 and 250 °C, YL fractions increase to 77 % and 80 % along with improvement in particle sizes to 2.66 nm and 3.1 nm, respectively. Mean while contribution of the BGL reduces to 23 % and 20 % for 2.66 and 3.1 nm NPs, respectively. This variation infers that $O^P$ vacancy defects are annihilated faster than the $O^B$ defects at these temperatures (TABLE I, Fig. S1).

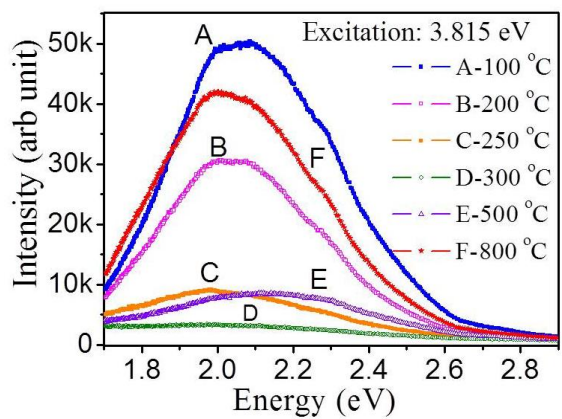

Fig. 6 PL of $SnO_2$ NPs annealed at different temperatures with excitation energy 3.815 eV (325 nm).

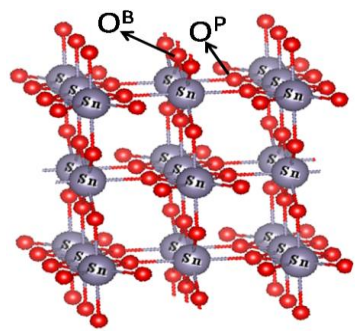

Fig. 7 Rutile tetragonal structure of $SnO_2$. $O^P$: in-plane oxygen, $O^B$: Bridging oxygen.

TABLE I. Gaussian deconvolution parameters of the spectra shown in Fig. 2.

| | | | | | Total YL | | | Total BGL |
|---|---|---|---|---|---|---|---|---|
| A (100 °C) | Position | 1.83 | 1.99 | 2.14 | | 2.29 | 2.45 | |
| | Area | 2,788 | 9,720 | 8,243 | 20,751 | 5,959 | 1,544 | 7503 |
| | % | 9.8% | 34.4% | 29.1% | 73.3% | 21.3% | 5.4% | 26.7% |
| B (200 °C) | Position | 1.81 | 1.97 | 2.12 | | 2.29 | 2.47 | |
| | Area | 2,607 | 4,524 | 4,777 | 11,908 | 2,581 | 959 | 3,540 |
| | % | 16.8% | 29.2% | 30.9% | 77% | 16.7% | 6.2% | 23% |
| C (250 °C) | Position | 1.77 | 1.96 | 2.13 | | 2.29 | 2.46 | |
| | Area | 1,015 | 1,550 | 1,060 | 3,625 | 556 | 314 | 870 |
| | % | 22% | 34.5% | 23.6% | 80% | 12.3% | 7% | 20% |
| E (500 °C) | Position | 1.77 | 1.97 | 2.13 | | 2.29 | 2.46 | |
| | Area | 457 | 979 | 1,167 | 2,603 | 896 | 387 | 1,283 |
| | % | 12% | 25.2% | 30% | 67% | 23% | 10% | 33% |
| F (800 °C) | Position | 1.81 | 1.96 | 2.11 | | 2.26 | 2.45 | |
| | Area | 3,228 | 5,418 | 5,485 | 14,131 | 4,407 | 1,251 | 5,658 |
| | % | 16.3% | 27.3% | 27.7% | 71% | 22.2% | 6.3% | 29% |

Sample D (300 °C) does not show any significant PL intensity around 2 eV (Fig. 6). In sample E (500 °C) of particle size 9 nm YL fraction decreases to 67% while BGL increases to 33 % (TABLE I). With respect to sample A, $O^P$ vacancy fraction is found to be higher than that of the $O^B$ defects for the sample E. The YL and BGL fractions change to 71 % and 29 % for sample F (800 °C) of particle size 25 nm (TABLE I). It is understandable from the above analysis that a dominant oxidation process takes place at low temperature of 200 °C unlike the oxidation at high temperature for thin film sample.[37] Rate of annihilation process of $O^P$ is found to be high in comparison to the $O^B$ vacancies (TABLE I). At high annealing temperature of 800 °C, the $O^B$ vacancies are found to dominate over $O^P$ vacancies. It is understandable that significant grain growth started at 500 °C as discussed in the previous section. At this temperature small grains will come together to form a big particle in order to reduce free energy of the system. In this process of growth from 2.4 nm to 9.5 nm, $O^P$ vacancies are created as effectively as $O^B$ vacancies. At annealing temperature of 800 °C, total PL intensity increases strongly indicating occurrence of large defects on the surface of NPs. In this sample $O^P$ vacancy percentage is much less than that of the $O^B$ vacancy percentage. It is noteworthy that the vacancy creation energy is higher for $O^P$ than that of $O^B$.[38] In summary, 'O' vacancies ($O^B$ and $O^P$) are very high in the sample A of size 2.4 nm. However, major contribution to PL intensity is from (73.3%) from $O^B$ vacancies. After annealing the sample A at 200 and 250 °C the NPs sizes increase to 2.66 and 3.1 nm, respectively and the overall 'O' vacancies are found to decrease. At this low temperature annealing process $O^B$ vacancy percentage increases over $O^P$ vacancy percentage (TABLE 1). Oxygen defects are considerably low for the NPs of size of 4 nm, which is obtained by annealing the sample A at 300 °C. Similarly annealed sample at 500 °C (9 nm), the PL intensity starts increasing. Here $O^P$ vacancy percentage also increases. Interestingly it has the highest percentage among all the samples (TABLE 1). Both 'O' vacancies ($O^B$ and $O^P$) increases further for the 25 nm NPs over the 9 nm $SnO_2$ NPs. However, $O^P$ vacancy percentage is found to decrease.

**Photoluminescence of –OH group related defects**
PL studies of $SnO_2$ NPs are also conducted using low energy excitations of 2.42 eV and 2.54 eV. Fig. 8 shows the PL spectra with the excitation energy of 2.42 eV. At a glance, the peak around 1.96 eV appears due to luminescence from the NPs. However after testing with another excitation of 2.54 eV (Fig. 9), it turns out to be a Raman mode related to -OH vibration (~ 3400 cm$^{-1}$). The observed Raman shift is consistent with respect to both the excitations and matches to the reported data for hydroxyl group as typical Sn-OH vibration mode.[39,40] In support to the presence of –OH group, FTIR measurements are also conducted typically on the samples C (250 °C) and F (800 °C). It confirms the strong presence of –OH groups in the sample C (Fig. 10). Peaks at 587, 1640 and 3400 cm$^{-1}$ are ascribed to various Sn-OH vibrations.[39] Strong presence of these peaks in the sample C (250 °C) over the sample F (800 °C) indicates the presence of high amount of –OH groups in low annealing temperature samples. Peak at 624 cm$^{-1}$ is assigned to the bulk O-Sn-O vibrational mode and is found to be strong in the sample F (800 °C) relative to the sample C (250 °C). Defect related peaks at 2.34 eV corresponding to Raman mode at 575 cm$^{-1}$ (D mode) and 2.33 eV ($A_{1g}$ mode at 632 cm$^{-1}$) are also seen in Fig. 8. These Raman modes are discussed (Fig. 4) previously. Apart from these Raman modes there is an interesting luminescence background around 1.96 eV. The same luminescence is observed with an excitation of 2.54 eV also (Fig. 9). In contrast to the PL intensity observed in Fig. 6, intensities in Figs. 8 and 9 are increased with annealing temperature and then it decreases for high annealing temperatures. Highest luminescence intensity is observed for the sample D (300 °C) with a 2.42 eV excitation energy (Fig. 8).

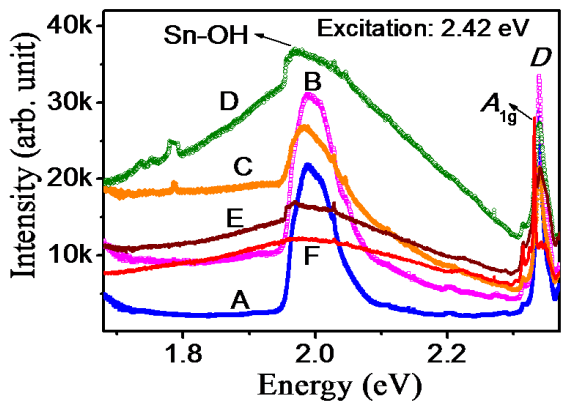

Fig. 8. Photoluminescence of SnO$_2$ NPs annealed at different temperatures with excitation energy 2.42 eV (514.5 nm).

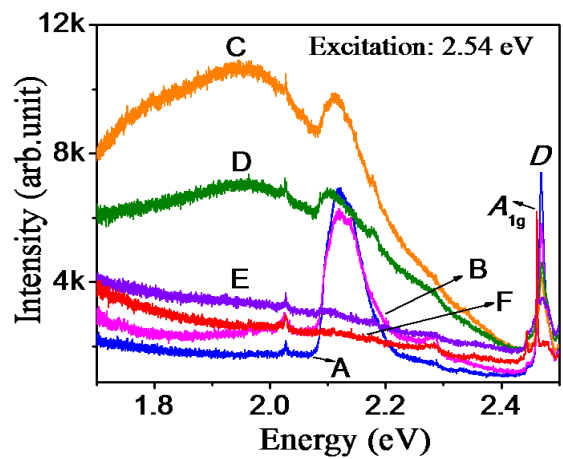

Fig. 9 Photoluminescence of SnO2 NPs annealed at different temperatures with excitation energy 2.54 eV (488 nm).

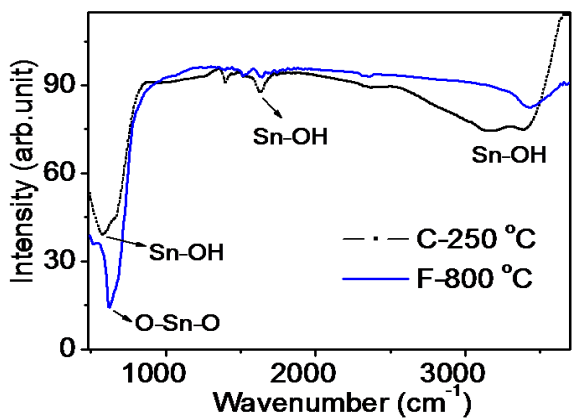

Fig.10 FTIR of SnO$_2$ NPs.

Whereas 2.54 eV excitation offers highest luminescence intensity (Fig. 9) for the sample C (250 °C) among all NPs. Samples A (100 °C) and B (200 °C) have nearly no luminescence intensity while the sample F (800 °C) has considerably low PL intensity with the 2.42 eV excitation (Fig. 8). All these three samples emit large luminescence intensity with the excitation of 3.815 eV (Fig. 6). Sample D (300 °C) shows hardly any luminescence with the 3.815 eV excitation (Fig. 6), whereas high PL intensity results for the same sample with low energy excitations of 2.54 eV and 2.42 eV (Figs. 8 and 9). From the above observations, the origin of the luminescence with low excitation energies (Figs. 8 and 9) strongly differs from the origin of luminescence with the high excitation energy (Fig. 6). Hence the luminescence observed in the low energy excitation may not be related to the 'O' vacancies. It was reported that the –OH groups on oxides were acted as luminescence centres. Luminescence around 2.1 eV was observed in ZnO NPs due to the presence of -OH groups.[22] R. G. Pavelko et al. showed using time-resolved diffuse reflectance in situ FTIR spectroscopy (DRIFTS) and mass spectrometry (MS) analysis that the -OH groups formation at bridging 'O' site of SnO$_2$ induced huge number of 'O' vacancies which acted as major donor states. Moreover, these bridging hydroxyl groups and surface oxygen closely interacted and participated in the creation of surface donor states. These donor states reduced the resistance of the material.[41] Based on these observations we may attribute the observed luminescence with low energy excitation to -OH groups formed on the SnO$_2$ NPs.

Samples A (100 °C) and B (200 °C) (Fig. 8) have high intensity of Sn–OH Raman vibrational mode. However, these two samples do not show any luminescence for low energy excitations (Figs. 8 and 9). From the literature it is clear that the conduction of oxides increases after formation of –OH groups on surface and it forms donor states close to the conduction band.[18,23,24,41] Optical band gap of the samples A and B are found to be around 4.45 and 4.25 eV, respectively (Fig. 5). Donor states created by –OH groups in these two samples are close to the conduction band and the excitation energies 2.54 eV and 2.42 eV are not sufficient to excite the electrons from valence band to –OH group related defect energy state. However, highest luminescence intensities are shown by the sample C (250 °C) and D (300 °C), having low band gaps around 4.06 and 3.90 eV, with the low energy excitations of 2.54 and 2.42 eV respectively. Noteworthy, luminescence related to –OH is not observed for the same sample D (300 °C) with the excitation energy of 3.815 eV (Fig. 6). This is ascribed to the heat energy due to the absorption of 3.815 eV which in turn destroys the –OH groups. Importantly, we do not observe any –OH related Raman peak with 325 nm (3.815 eV) laser for the samples C, D, E and F (supplementary Fig. S2). This observation supports instability of –OH functional group with the high excitation energy.

To get a better understanding on luminescence of –OH defect states, we have performed temperature dependent PL study. Sample B (200 °C) is chosen for this study with an excitation energy of 2.42 eV. There is hardly any luminescence at room temperature (Fig. 11), as 2.42 eV is not sufficient for exciting the –OH defect states in sample B, as discussed earlier. Interestingly luminescence is not observed at temperatures of 50, 100 and 150 °C also. Luminescence appears at the temperature of 200 °C. It is noteworthy that the band gap of a semiconductor decreases as the temperature increases.[42] In the present study, until a temperature of 200 °C, sufficient energy to the electrons is not provided. In consequence, electrons could not reach to the corresponding defect energy state. After recording PL at 200 °C, the material is cooled down to 25 °C and subsequent PL spectrum is now recorded. The corresponding PL intensity decreases and it almost matches to the room temperature PL intensity. The corresponding Raman peak shape for –OH group appears again as seen before exposing to the heating. It explains that –OH related defect band with 2.42 eV can be reached only after providing a minimum temperature of 200 °C. The energy supplied by 200 °C may be low, however, the subsequent reduction in band gap[42] and possible two step excitation process[43] might have helped in exciting the electron to the –OH group related defect state. This experiment offers further support for the defect band as presumed in samples A (100 °C) and B (200 °C) which cannot be attained using low energy excitations of 2.54 and 2.42 eV. Sn-OH vibrational mode recovers well after cooling down to room temperature by absorbing moisture from experimental ambient conditions (Fig. 11). This feature can be utilized in understanding the presence of the moist environment.

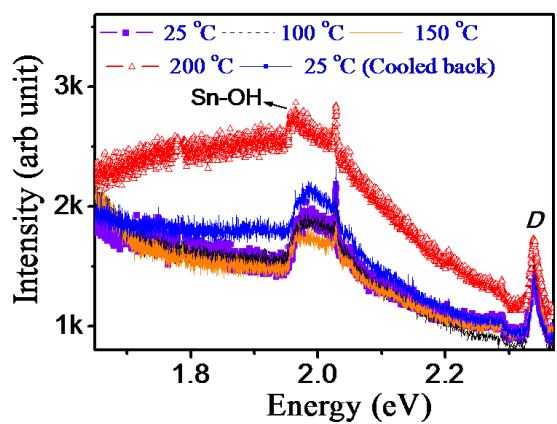

Fig. 11 Temperature dependent PL study of sample B (200 °C) with excitation energy of 2.42 eV.

## Conclusion

Photoluminescence intensities of $SnO_2$ QDs around 2 eV was correlated to various 'O' related defects which was found to decrease with increasing annealing temperature up to 300 °C. Further increase in the luminescence intensity with increasing annealing temperature was detected for the large NPs and was accounted for defects arising due to desorption of oxygen occurred on the surface of NPs at high temperatures. In addition to oxygen vacancy related defects, luminescence with low energy excitations from the –OH group defects is first time elucidated for the $SnO_2$ NPs. It is well known that the defects in $SnO_2$ play a crucial role in its powerful applications. We believe that the information provided here related to 'O' vacancies and -OH group functionalization of $SnO_2$ will help greatly to understand the nature of defects, so that one can manipulate them for efficient use of $SnO_2$ in technologically important applications.

## Acknowledgement

We thank Dr. Manas Sardar, MPD, MSG, IGCAR for the valuable discussions.

## Notes and references

[a]Surface and Nanoscience Division, [b]Materials Physics Division, Indira Gandhi Center for Atomic Research, Kalpakkam-603102, India. E-mail: dasa@igcar.gov.in, ramana9hcu@gmail.com

† Electronic Supplementary Information (ESI) available: [Table .S1, Gaussian fitting of PL spectra in Fig. 6 (Fig.S1) and Raman spectra of NPs with 325 nm (Fig. S2)]. See DOI: 10.1039/b000000x/

**Supplementary Information:**

**TABLE SI.** Size and Band gap of the SnO$_2$ NPs with respect to annealing temperature.

| Sample (Annealing Temperature) | Calculated size Using LFRS (nm) | Measured size using TEM (nm) | Band gap (eV) |
|---|---|---|---|
| A (100 °C) | 2.41±0.15 | 2.4±0.1 | 4.4 ± 0.05 |
| B (200 °C) | 2.66±0.18 | - | 4.25 ± 0.05 |
| C (250 °C) | 3.1±0.22 | - | 4.06 ± 0.05 |
| D (300 °C) | 3.7±0.27 | 4.1±0.2 | 3.9 ± 0.05 |
| E (500 °C) | - | 9.5±0.7 | 3.67 ± 0.05 |
| F (800 °C) | - | 25±1 | 3.62 |

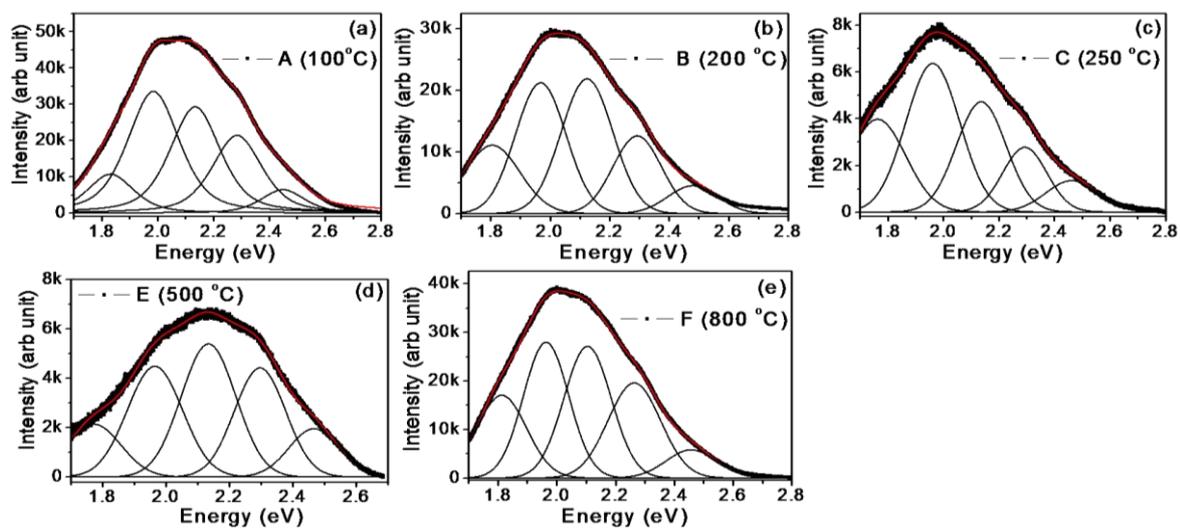

**Fig. S1.** Gaussian deconvolution of the PL spectra shown in Fig. 6. Area under each curve is deduced and shown in Table 1

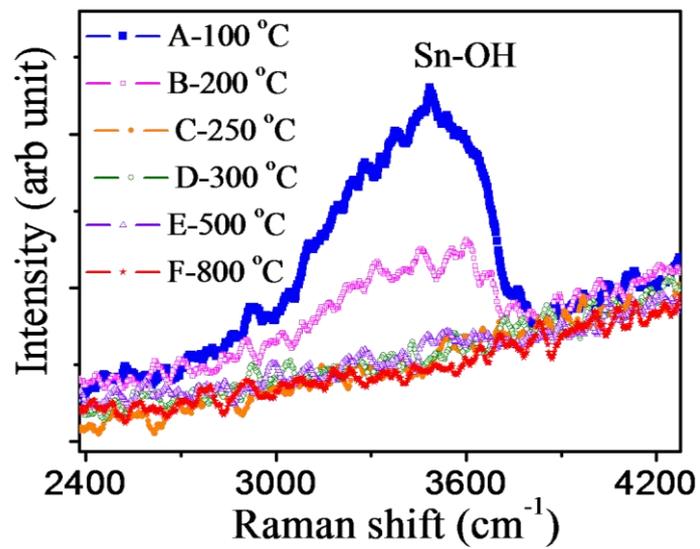

**Fig. S2**. Hydroxyl Raman mode of different size $SnO_2$ NPs with excitation 325 nm.